\documentstyle[12pt,aasms4]{article}

\begin{document}

\title{Are Abell Clusters Correlated with Gamma-Ray Bursts?}
\author{K. Hurley}
\affil{University of California, Space Sciences Laboratory, Berkeley, CA 94720-7450}
\authoremail{khurley@sunspot.ssl.berkeley.edu}

\author{D. Hartmann}
\affil{Clemson University, Dept. of Physics and Astronomy, Clemson SC 29634-1911}

\author{C. Kouveliotou, G. Fishman} 
\affil{NASA Marshall Space Flight Center, ES-62, Huntsville AL 35812}

\author{J. Laros}
\affil{Lunar and Planetary Laboratory, University of Arizona, Tucson, AZ 85721}

\author{T. Cline}
\affil{NASA Goddard Space Flight Center, Code 661, Greenbelt, MD 20771} 

\author{M. Boer}
\affil{CESR, 9, avenue du Colonel Roche, 31029 Toulouse, France}
 
\begin{abstract}

A recent study has presented marginal statistical evidence that
gamma-ray burst sources are correlated with Abell clusters, based on
analyses of bursts in the BATSE 3B catalog. Using 
precise localization information from the 3rd Interplanetary Network,
we have reanalyzed this possible correlation. We find that most of
the Abell clusters which are in the relatively large 3B error circles 
are not in the much smaller IPN/BATSE error regions.  We believe that 
this argues strongly against an Abell cluster--gamma--ray burst correlation.

\end{abstract}


\section{Introduction}

A correlation between the positions of gamma--ray bursts (GRBs) determined
by the Burst and Transient Source Experiment (BATSE) aboard the Compton
Gamma--Ray Observatory  and those of
rich, nearby Abell clusters was recently claimed by  Kolatt and Piran (\markcite{kp96} 1996) (hereafter KP).  They analyzed
the 
BATSE 3B catalog (Meegan et al. \markcite{me96} 1996) in conjunction
with data on Abell clusters (Abell, Corwin, $\&$ Olowin \markcite{aco89}
1989), and concluded that these gamma-ray bursts
were correlated with them at the 95\% confidence level. In
their study, they selected the 3B bursts with error circle radii 
$\lesssim$ 2.77 \arcdeg, and bursts and clusters with $\mid b \mid >
30 \arcdeg$. They then
calculated the number of burst--cluster pairs, N($\theta$), whose
separation was smaller than a given angle $\theta$, for $\theta$=1\arcdeg 
-- 6\arcdeg. Comparing this number to the numbers found for
randomly generated catalogs, they found a number for $\theta$=
4\arcdeg that was significant at the 95\% confidence level. 

If such findings could be confirmed,
they would constitute statistical evidence for counterparts to
GRB sources, and would indicate that at least some GRBs are at cosmological
distances.  We have attempted to confirm these results by subjecting them to more
stringent tests. The positions of those Abell clusters which appear to be 
related to GRBs, based on their locations within BATSE error circles, must
be consistent with {\it all} known localization information for the bursts
in question. In particular, they must also lie within the annuli or the
error boxes of the 3rd Interplanetary Network (IPN3) for those bursts.
Below we present the IPN3 data, the results of our test, and our conclusions.  

\section{IPN3 Data}

For the period covered by the 3B catalog, IPN3 consisted of the Ulysses spacecraft,
 at distances up to 6 AU
(Hurley et al. \markcite{hu92} 1992),
BATSE, and, until mid-1992, Pioneer Venus Orbiter (PVO). When only Ulysses and BATSE
observed a burst, the resulting localization was an annulus generally crossing 
the BATSE error circle.
When PVO observed the burst also, an error box resulted with dimensions
in the several arcminute range in the best cases. Examples
may be found in Hurley et al. (\markcite{hu93} 1993) and Cline et al.
(\markcite{cl92} 1992). The annuli widths varied from about 10" to 1000", with
average width 5.7'. Descriptions of the data set have appeared in Hurley
et al. (\markcite{hu94} 1994, \markcite{hu96} 1996). Of the 136 ``accurately''
localized BATSE bursts selected by KP on the basis
of their error circle size and galactic latitude, 93 have IPN3 annuli associated
with them, and 10 have IPN3 error boxes. The total area covered by the 10 error boxes, plus the 93 intersections of
the IPN3 and BATSE locations, is less than the area covered by the corresponding BATSE error
circles alone, by at least an order of magnitude. This substantially reduces the 
probability of chance correlations between
Abell clusters and GRB positions, and results in a stronger test.

\section{Results of the Test}

The all-sky catalog of Abell clusters (Abell, Corwin, $\&$ Olowin \markcite
{aco89} 1989) contains 4073 rich clusters, each having at least 30 bright members, and
covers redshifts less than z = 0.2. Following KP, a latitude
cut of $\vert{\rm b}\vert \ge$ 30$^{\rm o}$ was applied to both cluster and
burst catalogs.
We began by verifying the numbers of bursts and Abell clusters in the KP 
study, 136 (after latitude and error circle size cuts are applied) and 3616 
(after the latitude cut), respectively. For each cluster, we then checked each of
the BATSE error circles to see whether their positions were consistent. If it was,
we finally checked for consistency with any IPN3 location information. Since, in the KP study, Abell clusters within 4 $\arcdeg$ of a BATSE position were considered to
be correlated with the burst, we assigned an error radius of 4 $\arcdeg$ to the
BATSE bursts, and also used this to define the BATSE/IPN3 error box.  An example
is shown in figure 1.

We found that 1260 Abell clusters lie in the BATSE 4 $\arcdeg$ radius error circles.
These circles cover about 16.6\% of the sky, and given the density of Abell clusters
in the sample, about 1200 would be expected to fall in the error circles by chance.
This is consistent with the KP result in that it represents about a 2 $\sigma$
excess of Abell clusters in these error circles, assuming a negligible cluster-
cluster correlation. The BATSE/IPN
error boxes cover 3.3x10$^5$ square arcminutes, or 1.9\% of the total area of the BATSE
error circles that they intersect. We found that 14 Abell clusters lie within the
BATSE/IPN3 error boxes, while 16 are expected by chance. This indicates that the overwhelming
majority of the Abell clusters which are in BATSE error circles are there by chance.

We reanalyzed the data using the standard 1 $\sigma$ BATSE error circle radii, defined as
$(\sigma_{sys}^2 + \sigma_{stat}^2)^{1/2}$, where $\sigma_{sys}$ is the systematic
error, 1.6 $\arcdeg$, and $\sigma_{stat}$ is the statistical error, given in the BATSE 3B
catalog for each burst. This results in smaller error circles, covering 3.5\% of the sky.
The BATSE/IPN3 error boxes cover 4.5\% of the total area of the BATSE error circles that
they intersect, or 1.2x10$^5$ square arcminutes, 
and 5.7 Abell clusters should fall in them by chance. 4 were found to
actually lie in them. 

\section{Discussion}

If gamma-ray bursts indeed originate at cosmological distances, their
observed brightness distribution suggests that BATSE samples bursts to
a redshift of order unity (e.g., Wickramasinghe et al. \markcite{w93} 
1993). To yield the all-sky rate of $\sim$ 10$^3$
yr$^{-1}$ an average galaxy must produce one observable burst every 10$^6$
yrs, corresponding to a mean comoving galaxy density of
$n_0 = 10^{-2}\ {\rm Mpc}^{-3}.$ In the local part of the universe
approximate distances are given by D = z L$_{\rm H}$, where the
Hubble distance is defined as 
${\rm L_H} = \rm{c}/{\rm H}_0 \sim 3,000 \ h^{-1} {\rm Mpc}$, where
h is the Hubble constant in units of 100 km s$^{-1}$ Mpc$^{-1}$.
The depth of the Abell cluster sample is thus about 600 h$^{-1}$ Mpc.

Only $\sim$ 2\% of all observable GRBs are thus expected to occur
within the volume of space sampled by the Abell catalog. 
Even if we assume that all galaxies are correlated with clusters,
to obtain a significant correlation of bursts with clusters would thus
require that either the BATSE sampling redshift is much less than unity,
and/or that all well--localized bursts are approximately confined to 
the sampling volume of the cluster catalog. KP
obtain a sample of 549 bursts after applying their first cut
(in latitude).
We expect $\sim$ 2 \% of these (11) to be correlated with clusters.
As shown above, concidences
are expected in 5.7 cases, and the observed number of coincidences is 4. 

The KP results indicate a sampling distance 
(for all bursts)
of z$_{\rm max}$ = 0.7 and a sampling distance of z$_{\rm a}$ = 0.31 for
the 136 ``accurately'' localized bursts, which implies that 35 ($\pm$ 20)
bursts are within the Abell range. Thus, in addition to the number of chance
coincidences between Abell clusters and the IPN3 positions (5.7), one expects 15$-$55 
more, which is clearly in contradiction with our findings.

We believe that the Abell cluster--gamma--ray burst correlation found by KP
is best explained by statistics. There is a 5\% chance of finding a correlation 
with any set of cataloged objects in their study,
and certainly many catalogs have been searched for GRB counterparts. Indeed, as soon as
precise GRB positions began to become available from the 1st IPN, they were subjected to
extensive catalog searches, including Abell clusters, (e.g. Hurley \markcite{hu82} 1982, Barat et al. \markcite
{ba84} 1984) and it seems unlikely that an Abell cluster correlation would have been
missed. 

If a correlation were to exist between Abell clusters and GRBs, the poor
angular resolution of BATSE, the small sample size for the IPN3 positions, and
the small number of low redshift events reduce our chances
of detecting it. However, Abell clusters are somewhat
concentrated towards the super-galactic plane to distances perhaps as large
as $\sim$ 300\ h$^{-1}$ Mpc
(e.g., Tully 1987 \markcite{tu87}; but see Postman {\it et al.} 1989 \markcite{post89}) 
and one might thus
expect to find a significant global anisotropy of bright bursts tracing the galaxies
within z $\sim$ 0.1. This effect could be noticeable even if only poor localizations were 
available. So far, the search for a super-galactic anisotropy has not been successful 
(Hartmann et al. 1996). 

While attempts to find GRB counterparts through
correlation analysis with galaxies or clusters of galaxies may provide some evidence
for a cosmological burst origin, we caution that the present data are insufficient 
to prove that connection. Five more years of BATSE/IPN3 data may be sufficient for this 
task.  Another solution would be to improve burst localization accuracy to
the arcsecond regime, so that one would not have to rely on statistics, but simply on
the inspection of such error boxes. This may become possible with ongoing searches for
optical transients, through future imaging burst detectors,
or an improved triangulation network.  All of these possibilities are being
pursued actively at the present time.

\acknowledgments

KH is grateful for Ulysses support under JPL Contract 958056, and IPN3 support under
NASA NAG 5-1560. DH acknowledges support from NASA through the COMPTON Observatory
guest investigator program.  We also acknowledge the helpful comments of Chip Meegan,
and the careful reading and constructive criticism of T. Kolatt and T. Piran.

\begin{figure}
\caption{The field around BATSE burst 121.  The BATSE 4 \arcdeg  radius
error circle and the Ulysses/BATSE triangulation annulus are shown. None
of the Abell clusters in this field lies within the triangulation annulus.}
\end{figure}

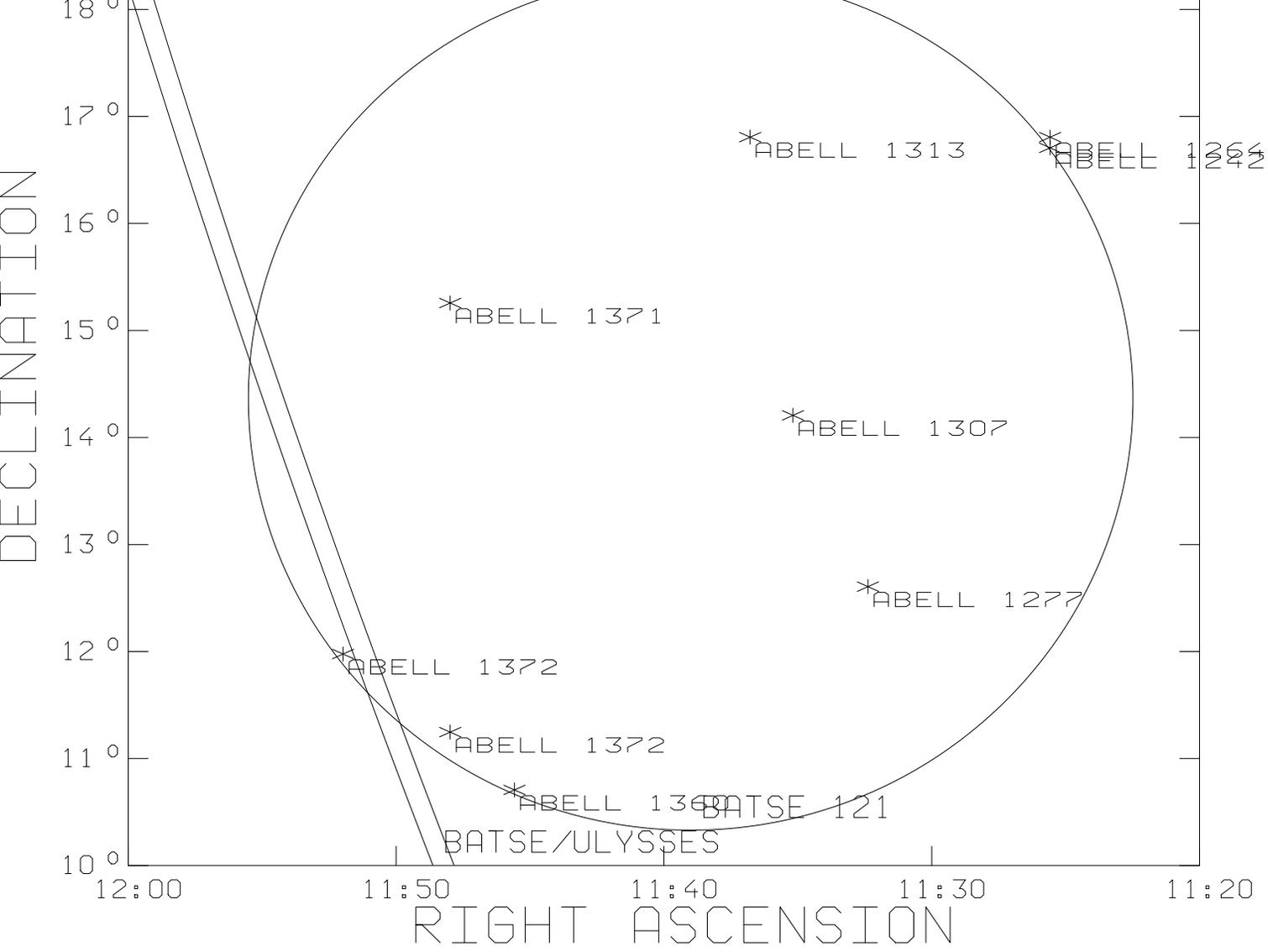
\end{document}